\documentclass[journal]{IEEEtran}

\usepackage{amsthm,amsfonts,amssymb,mathtools,nccmath,mathrsfs}
\usepackage[lined,linesnumbered,ruled,commentsnumbered]{algorithm2e}
\usepackage{pgfplots}
\usepackage{xcolor}
\usepackage{microtype}
\usepackage{url}
\usepackage{setspace}
\usepackage{array}
\usepackage{subfigure}
\usepackage{textcomp}
\usepackage{stfloats}
\usepackage{verbatim}
\usepackage{graphicx}
\usepackage{cite}
\usepackage{hyperref}
\usepackage{soul}
\usepackage{tikz}
\usetikzlibrary{arrows,shapes,backgrounds,plotmarks,positioning}
\usepackage{interval}
\usetikzlibrary{decorations.pathreplacing}
\pgfplotsset{compat=newest} 
\pgfplotsset{plot coordinates/math parser=false}
 \newlength\figureheight
\newlength\figurewidth

\newtheorem{definition}{Definition}

\newtheorem{proposition}{Proposition}

\newtheorem{remark}{Remark}

\newcommand{\X}{\mathcal{X}}

\newcommand{\Y}{\mathcal{Y}}
\newcommand{\Sen}{\mathcal{S}}

\newcommand{\E}{\mathbb{E}}

\newcommand{\PX}{P_{X}}
\newcommand{\PS}{P_{S}}
\newcommand{\PY}{P_{Y}}

\newcommand{\PXY}{P_{XY}}

\newcommand{\PXgy}{P_{X|Y=y}}

\newcommand{\PBS}{{\mathbf{P}}_{S}}
\newcommand{\PBX}{{\mathbf{P}}_{X}}
\newcommand{\PBY}{\mathbf{P}_{Y}}

\newcommand{\PBSgx}{\mathbf{P}_{S|X=x}}

\newcommand{\PBSgy}{\mathbf{P}_{S|Y=y}}

\newcommand{\PBXgy}{\mathbf{P}_{X|Y=y}}

\newcommand{\PBSX}{\mathbf{P}_{SX}}
\newcommand{\PBSY}{\mathbf{P}_{SY}}
\newcommand{\PBXY}{\mathbf{P}_{XY}}
\newcommand{\PBSgX}{\mathbf{P}_{S|X}}

\newcommand{\PBXgY}{\mathbf{P}_{X|Y}}
\newcommand{\PBYgX}{\mathbf{P}_{Y|X}}

\newcommand{\e}{\mathrm{e}}
\newcommand{\eps}{\varepsilon}

\newcommand{\Real}{\mathbb{R}}
\newcommand{\Dalpha}[2]{D_{\alpha}\left(#1\|#2\right)}
\newcommand{\ellalpha}[2]{\ell_{\alpha}\left(#1\|#2\right)}

\DeclareMathOperator*{\argmax}{arg\,max}

\begin{document}

\title{Privacy-Utility Tradeoff Based on $\alpha$-lift }

\author{
    Mohammad A.~Zarrabian,~\IEEEmembership{Member,~IEEE,} and~Parastoo~Sadeghi,~\IEEEmembership{Senior Member,~IEEE}
    \thanks{Mohammad A.~Zarrabian is with the College of Engineering, Computing, and Cybernetics, Australian National University, Canberra, Australia, e-mail: mohammad.zarrabian@anu.edu.au. Parastoo Sadeghi is with the School of Engineering and Technology, the University of New South Wales, Canberra, Australia, e-mail: p.sadeghi@unsw.edu.au.}
}

\maketitle

\begin{abstract}
    Information density and its exponential form, known as lift, play a central role in information privacy leakage measures. $\alpha$-lift is the power-mean of lift, which is tunable between the worst-case measure max-lift ($\alpha=\infty$) and more relaxed versions ($\alpha<\infty$).
    This paper investigates the optimization problem of the privacy-utility tradeoff (PUT) where $\alpha$-lift and mutual information are privacy and utility measures, respectively. 
    Due to the nonlinear nature of $\alpha$-lift for $\alpha<\infty$, finding the optimal solution is challenging. 
    Therefore, we propose a heuristic algorithm to estimate the optimal utility for each value of $\alpha$, inspired by the optimal solution for $\alpha=\infty$ and the convexity of $\alpha$-lift with respect to the lift, which we prove. The numerical results show the efficacy of the algorithm and indicate the effective range of $\alpha$ and privacy budget $\eps$ with good PUT performance.
\end{abstract}

\section{Introduction}

    In our data-driven world, data sharing with third parties poses significant privacy risks, especially given recent advances in machine learning and signal processing. To protect sensitive information, data is often randomized by a privacy mechanism before sharing. The design of these mechanisms depends on the measures that quantify privacy leakage and data usefulness, leading to the privacy-utility tradeoff (PUT) problem.
    
    Among privacy measures, differential privacy (DP) \cite{2006CalibNoiseDFP,2006DFP} and local differential privacy (LDP) \cite{2003LimitPrivBreach,2011Learnprivately,2014ExtermalMechanism,2014LDPrateDist,2017extremepointLDP} are dominant. 
    However, DP and LDP are independent of the prior data distribution and can lead to low data utility \cite{saeidian2023inferential,2021ContextawareLIP}.   
    In contrast, information-theoretic measures quantify privacy risk by information leakage between the randomized and sensitive data. 
    They can be prior-dependent and are also known as context-aware privacy in the literature. 
    These measures include local information privacy (LIP) \cite{2012PrivStatisticInfere,2020LIPDataAggr,2021ContextawareLIP,2023Onthelift,grosse2024quantifying}, maximal leakage \cite{2020MaxL,saeidian2021optimalmaximalleakage}, pointwise maximal leakage (PML) \cite{2022PML,saeidian2023pointwise,2023extremalPML}, mutual information, \cite{2013UPTdatasets,2012PrivStatisticInfere} and (maximal) $\alpha$-leakage \cite{2019TunMsurInfLeak_PUT,2019Robustness-alpha-leak,2020alphaprperties}.
    
    Privacy measures could be further classified into average and pointwise measures.  
    Average measures quantify the expected privacy leakage with respect to (w.r.t) the marginal distribution of the randomized data. These include 
    maximal leakage, mutual information, and $\alpha$-leakage.
    However, it has been argued that average measures do not guarantee privacy for each individual realization of the randomized data \cite{2012Alvimgain,2012PrivStatisticInfere,2021ContextawareLIP,2022PML}.  
    Therefore, pointwise measures such as  LIP and PML have been proposed to achieve this goal. 
    
    The maximal $\alpha$-leakage \cite{2019TunMsurInfLeak_PUT}, $\alpha \in [1,\infty]$, which is equivalent to the Sibson mutual information \cite{1969SibsonInfRad}, offers a tunable privacy measure that can provide varying levels of privacy based on the value of $\alpha$. It is equal to the mutual information when $\alpha=1$, and as $\alpha$ increases, it becomes a more stringent measure. For $\alpha=\infty$, it becomes maximal leakage, which is the worst-case average measure and is also known as Bayes capacity~\cite{quantitiveInformationFlow}.
    While $\alpha$-leakage has not been extended to the pointwise version in a strict sense, a closely related measure has been introduced in \cite{2021Alpha-LiftWatchdog}, named $\alpha$-lift ($\alpha \in (1,\infty]$).\footnote{$\alpha$-lift is a \emph{semi-pointwise} measure: it is a pointwise measure w.r.t the mechanism output, but is an average measure w.r.t the sensitive/input data. In the same vein, strong $\chi^2$~\cite{2021StrongChi2}, and $\ell_{1}$-norm~\cite{2022DataDsclsurell1Priv} are semi-pointwise measures.}  It is closely related to the R{\'e}nyi divergence \cite{1961measuresreny} and Sibson mutual information \cite{1969SibsonInfRad}.
    $\alpha$-lift is a power-mean or $\alpha$-norm of a quantity known as \emph{lift} (see below). $\alpha$-lift  is tunable between worst-case maximum lift for $\alpha=\infty$ and gives more relaxed measures for $1 < \alpha < \infty$.
    
    Indeed, lift is central to many information-theoretic leakage measures, whose logarithm is known as information density and is the fundamental measure of information gain between two distributions \cite{1961measuresreny}. The operational meaning of maximum lift was studied in inferential privacy and guessing framework in the context of PML \cite{2022PML} and LIP \cite{2012PrivStatisticInfere}.

    Data usefulness is usually quantified by a statistical distance between original and randomized data, such as Hamming distortion \cite{2018DPHammDistr,zhong2022privacyutility} and minimum mean squared error \cite{2021ContextawareLIP}. Another widespread class of utility measures is sub-convex functions, including mutual information, Kullback-Leibler divergence, and $f$-divergence measures~\cite{2014ExtermalMechanism,2021StrongChi2,2021DataSanitize,2023Onthelift,2022DataDsclsurell1Priv}. 
    While the optimal mechanism for sub-convex functions has been obtained based on LDP, PML, and LIP, achieving this optimal solution is challenging for $\alpha$-tunable and some other measures due to their nonlinear nature. 
    Tight estimations of optimal mechanisms based on strong-$\chi^2$ and $\ell_1$-norm privacy were given for very high privacy regime in \cite{2021StrongChi2,2022DataDsclsurell1Priv}.

    In \cite{2021Alpha-LiftWatchdog}, a merging privatization based on the watchdog mechanism \cite{2019Watchdog,2020PropertiesWatchdog} was applied to achieve $\alpha$-lift privacy.  
    In \cite{2023Onthelift}, an improved watchdog method, named subset merging \cite{2022EnhanceUtilWatchdog}, was utilized to enhance utility measured by mutual information. However, none of these methods were aimed at optimizing PUT.
    To the best of our knowledge, the optimal PUT of sub-convex utility functions and $\alpha$-tunable  privacy measures have not been considered before.
    
    We study the optimal privacy-utility tradeoff for $\alpha$-lift, where utility is measured by mutual information as an instance of subconvex functions. Our contributions are as follows:

    \begin{itemize}
        \item We prove the convexity of $\alpha$-lift w.r.t the lift and apply this property to study the privacy utility tradeoff.
        \item Since finding the optimal solution is challenging due to the nonlinear nature of $\alpha$-lift for $\alpha < \infty$, we propose a heuristic algorithm to estimate the optimal utility. To our best knowledge, this algorithm is the first of its kind. 
        \item Finally, we evaluate the efficacy of the algorithm via numerical simulations and study the effect of different values of $\alpha$ and privacy budget on the utility values.
    \end{itemize}


    \subsection{Notation}

        We consider discrete random variables defined on finite alphabets. 
        Random variables are denoted by capital letters, e.g., $X$, and their realizations by lowercase letters, e.g., $x$. 
        Sets are represented by uppercase calligraphic letters, such as $\X$ for the alphabet of $X$ with cardinality $|\X|$.
        Vectors and matrices are shown by bold uppercase letters e.g., $\mathbf{V}=[V_{1},V_{2},\cdots,V_{n}]^{T}$.
        For random variables $X$ and $Y$, $\PBXY$ is the matrix of their joint probability distribution with $|\X|$ rows and $|\Y|$ columns and its elements are denoted by $\PXY(x,y)=\Pr[X=x,Y=y]$.
        The marginal distribution of $X$ is  denoted by $\PBX$, a vector with elements $\PX(x)=\Pr[X=x]$.
        The conditional probability of $X$ given a realization $Y=y$ is denoted by $\PBXgy$ with elements $\PXgy(x)=\Pr[X=x|Y=y]$.


\section{Preliminaries and System Model}

    Denote $X$ as the useful data intended to be shared, which is correlated with sensitive data $S$ via the joint distribution $\PBSX \neq \PBS\PBX^{T}$.
    To protect $S$, another random variable $Y$ will be published as the output of the privacy mechanism $\PBYgX$ with input $X$, where they form a Markov chain $S-X-Y$, resulting in $\PBSgy=\PBSgX\PBXgy,\forall y\in\Y$.
    Denote by $i(s,y)=\log\frac{p(s|y)}{p(s)}$, information density, the pointwise information gain \cite{1969SibsonInfRad,grosse2024quantifying} for the adversaries.
    The exponential value of $i(s,y)$ is the multiplicative adversarial gain known as lift~\cite{2019Watchdog,2020PropertiesWatchdog,2022EnhanceUtilWatchdog,2022asymmetric,2023Onthelift,2022EXplainEps}:$$l(s,y)=\frac{p(s|y)}{p(s)}.$$
    We now study the $\alpha$-lift privacy measure proposed in \cite{2021Alpha-LiftWatchdog}.
    \begin{definition}\label{def:alphalift}
        Given a joint distribution $\PBSY$ on finite alphabet $\Sen \times \Y$ and $\alpha \in (1,\infty]$, the $\alpha$-lift for each $y \in \Y$ is defined as
        \begin{align}\nonumber
            \ell_{\alpha}(\PBSgy\|\PBS)=
            \begin{cases}
                \displaystyle \left(\sum_{s\in \Sen} \PS(s) l^{\alpha}(s,y)\right)^{\frac{1}{\alpha}},  &\alpha \in (1,\infty), \\
                \displaystyle \max_{s \in \Sen}l(s,y), &\alpha=\infty. 
            \end{cases}
        \end{align}
    \end{definition}
  
    \begin{remark}\label{remark:non-decreasing}
        $\alpha$-lift is non-decreasing in $\alpha$, which is due to \cite[Thm. 3]{2014Renyikullbackprop} and noting the relation between the R{\'e}nyi divergence and $\alpha$-lift: $\Dalpha{\PBSgy}{\PBS}\!=\!\frac{\alpha}{\alpha-1}\log \ellalpha{\PBSgy}{\PBS}$.
    \end{remark}
    
    \begin{remark}\label{remark:alphainfity}
         By letting $\alpha=\infty$, Definition~\ref{def:alphalift} yields the max-lift, which is the upper bound in LIP \cite{2012PrivStatisticInfere,2019Watchdog}. When $S=X$ and $\alpha=\infty$, the entire useful data is treated as sensitive, and $\alpha$-lift coincides with PML \cite{2022PML}.
    \end{remark}
    
    Next, we show the convexity of  $\alpha$-lift w.r.t the lift, which will be used later for developing our heuristic algorithm.
    \begin{proposition}\label{prop:convex}
        $\alpha$-lift is convex w.r.t the lift $l(s,y)$. 
        \begin{IEEEproof}
            Let $\|f(S)\|_{\alpha}\!=\!\left(\E_{S}|f(S)|^{\alpha}\right)^{1/\alpha}$. Fix $y \in \Y$ and consider two distributions $\PBSgy'$ and $\PBSgy^{''}$, and let  $\PBSgy=\lambda\PBSgy^{'}+(1-\lambda)\PBSgy^{''}$. 
            Consequently,  $l(s,y)=\lambda l'(s,y)+(1-\lambda)l^{''}(s,y)$ and we have:
            \begin{align}
            \ell_{\alpha}(\PBSgy\|\PBS) 
            & =\|l(S,y)\|_{\alpha}\\
            &=\|\lambda l'(S,y)+(1-\lambda)l^{''}(S,y)\|_{\alpha}\\
            &\leq \|\lambda l'(S,y)\|_{\alpha}+\|(1-\lambda)l^{''}(S,y)\|_{\alpha}  \label{eq:Minkowski ineq} \\
            &=\lambda\|l'(S,y)\|_{\alpha}+(1-\lambda)\|l^{''}(S,y)\|_{\alpha},
            \end{align}
            where \eqref{eq:Minkowski ineq} is due to Minkowski inequality \cite{1952inequalities}.
        \end{IEEEproof}
    \end{proposition}
    
    \begin{definition}
        For $\eps \in \Real_{+}$ and $\alpha \in (1,\infty]$, a privacy mechanism $\mathcal{M}: \X\rightarrow \Y$ is $(\eps,\alpha)$-lift private w.r.t $S$, if:
        \begin{equation}
            \ellalpha{\PBSgy}{\PBS} \leq \e^{\eps},~~ \forall y \in \Y.
        \end{equation}
    \end{definition}
    Proposition~\ref{prop: epsmax} establishes an upper bound on $\alpha$-lift privacy.
    \begin{proposition}\label{prop: epsmax}
        All privacy mechanisms satisfy $(\eps^{\text{max}}_\alpha, \alpha)$-lift privacy, where $\eps^{\text{max}}_{\alpha}=\log \max_{x\in\X}\ellalpha{\PBSgx}{\PBS}$.
    \end{proposition}
    \begin{IEEEproof}
        For all privacy mechanisms, we have
        \begin{align}
            \nonumber &\max_{y\in \Y} \ellalpha{\PBSgy}{\PBS}
            = \max_{y\in \Y} \left(\sum_{s\in \Sen} \PS(s) l^{\alpha}(s,y)\right)^{\frac{1}{\alpha}}\\
            \nonumber  &= \max_{y\in \Y} \left(\sum_{s\in \Sen} \PS(s) \left(\sum_{x\in\X}\PXgy(x)l(s,x)\right)^{\alpha}\right)^{\frac{1}{\alpha}}\\
            &\label{eq:jensen}\leq  \max_{y\in \Y} \left( \sum_{s\in \Sen} \PS(s) \sum_{x\in\X} \PXgy(x) l^{\alpha}(s,x)\right)^{\frac{1}{\alpha}}\\
            &\label{eq:ineq replaces}= \max_{y\in \Y} \left( \sum_{x\in\X} \PXgy(x)\sum_{s\in \Sen} \PS(s) l^{\alpha}(s,x)\right)^{\frac{1}{\alpha}}\\
            \label{eq:maxapost} &=\max_{x}\left(\sum_{s\in \Sen} \PS(s) l^{\alpha}(s,x)\right)^{\frac{1}{\alpha}},
        \end{align}
        where \eqref{eq:jensen} is due to Jensen's inequality, \eqref{eq:ineq replaces} is given by replacement of summations
        and \ref{eq:maxapost} is given by maximum a posteriori decision, where $\PXgy=1$  for $y=\argmax_{x\in\X}\sum_{s\in\Sen}\PS(s)l^{\alpha}(s,x)$ and zero otherwise.
    \end{IEEEproof} 
 
    In the special case of $\alpha = \infty$, the PUT for the max-lift as the privacy measure and mutual information as the utility measure was studied in \cite{2021DataSanitize}.
    The optimization problem is 
    \begin{align}
        \label{eq:LIP optim} &\max_{\substack{\PBYgX}} I(X;Y) = H(X) - \min_{\substack{\PBXgY,\PBY}} H(X|Y)  \\ 
        \text{s.t.}~~ & \label{eq:maxlift cond} \PBSgX\PBXgy \preceq \e^{\eps}\PBS,~~ \forall y \in \Y,  \\ 
        & \label{eq:PXgY cond}\mathbf{1}^{T}\PBXgy=1,~\PXgy(x) \geq 0, \forall x,y \in \X\times\Y,   \\ 
        & \label{eq:Py cond}\mathbf{1}^{T}\PBY=1, ~\PY(y) \geq 0, \forall y \in \Y,  \\ 
        & \label{eq:Py&PXgy cond}\PBXgY\PBY=\PBX, 
    \end{align}
    where the symbol $\preceq$ represents the component-wise inequality between two vectors and $\mathbf{1}$ is all one vector.
    Since $H(X|Y)$ is concave w.r.t $\PBXgY$ and conditions \eqref{eq:maxlift cond}-\eqref{eq:PXgY cond} make a convex polytope, the feasible candidates for the optimal solutions $\PBXgy$ are among the vertices of this polytope. 
    Accordingly, the optimization is solved in the following steps:   
    \begin{enumerate}
        \item \label{step:1}
        Enumerate vertices of the following polytope as the candidates for the optimal $\PBXgy$:
        \begin{equation}\label{eq:polytope}
            \Pi_{\eps}=\left\{\hspace{-10pt}
            \begin{array}{ll}
                &\mathbf{V} \in \mathbb{R}^{|\X|}: \\
                &\mathbf{1}^{T}\mathbf{V
                }=1,~~  V_{x} \geq 0, \forall x \in \X,\\
                &\PBSgX\mathbf{V} \preceq \e^{\eps}\PBS.
            \end{array} \right\}
        \end{equation}
            
        \item \label{step:2}
            Let $\mathcal{V}_{\eps}=\{\mathbf{V}^{1}, \cdots, \mathbf{V}^M\}$ be the set of $\Pi_{\eps}$'s vertices and $h(\mathbf{V}^{i})$ the corresponding entropy of $\mathbf{V}^{i}$ as follows: 
            $$h({\mathbf{V}}^{i})=-\sum_{x\in \X}V_{x}^{i}\log V_{x}^{i}.$$
            Then $\PBXgY$ (with columns $\PBXgy$), and $\PBY$ are given by the solution of the following linear program:
            \begin{equation}\label{eq:Py optimize}
                \begin{aligned}
                \min_{\mathbf{Q} } & \sum_{i=1}^{M}Q_{i}h(\mathbf{V}^{i}),\\
                \text{s.t.}~~ &\mathbf{1}^{T}\mathbf{Q}=1,~Q_{i}\geq 0,\\
                \quad & \sum_{i=1}^{M}Q_{i}\mathbf{V}^{i}=\PBX .
                \end{aligned}
            \end{equation}  
        
        \item \label{step:3}
            Let $\mathcal{I}=\{i: Q_{i}>0 \}=\{\iota_{1},\cdots,\iota_{|\mathcal{I}|}\}$  be the set of indices of nonzero elements in $\mathbf{Q}$. Then the set $\Y$, $\PBY$ and $\PBXgY$ are given by  
        \begin{align}
            &\label{eq:output Y}\Y=\{1,\cdots,|\mathcal{I}|\},\\
            &\label{eq:output PY}\PY(y)=Q_{\iota_{y}},~ y \in \Y,\\
            &\label{eq:output Pxgy}\PBXgy=\mathbf{V}^{\iota_{y}},~ y \in \Y.
        \end{align}        
        \end{enumerate} 
        The main contribution of this paper, presented in the next two sections, is to extend this approach for optimizing PUT from $\alpha=\infty$ to the entire range $\alpha \in (1,\infty]$.
        
    
\section{Privacy-Utility Tradeoff based on $\alpha$-lift}
    
    Let us focus on the case where $\alpha \in (1,\infty)$. We introduce the optimization problem for the PUT of $\alpha$-lift as\footnote{Note that mutual information has been chosen for the sake of simplicity. Our proposed optimization framework and the heuristics algorithm are applicable to all sub-convex utility functions with $\alpha$-lift as the privacy measure.}
    \begin{align}\label{eq:alpha optim} 
        \max_{\substack{\PBYgX}} & I(X;Y) = H(X) - \min_{\substack{\PBXgY,\PBY}} H(X|Y), \\
        \text{s.t.}  ~~
        & \label{eq:alpha cond} \ellalpha{\PBSgy}{\PBS} \leq \e^{\eps},~\forall y \in \Y,
    \end{align}
    along with the same conditions \eqref{eq:PXgY cond}-\eqref{eq:Py&PXgy cond}.
    
    Since $l(s,y)=\sum_{x\in \X}l(s,x)\PXgy(x)$ for all $s\in \mathcal{S}$ and $\PBSgy=\PBSgX\PBXgy$, Proposition \ref{prop:convex} implies that $\ellalpha{\PBSgy}{\PBS}$ is convex also w.r.t $\PBXgy$. 
    Thus, the optimal solution for \eqref{eq:alpha optim} occurs when $\ellalpha{\PBSgy}{\PBS}=\e^{\eps}$. Let us write $\ellalpha{\PBSgy}{\PBS} = \ellalpha{\PBSgX \PBXgy}{\PBS}$.     
    For given $\alpha$ and $\eps$, denote by $\mathcal{F}_{\alpha}^{\eps}$ the set of feasible points $\PBXgy$ that satisfy the inequality in  \eqref{eq:alpha cond} and by $\mathcal{E}_{\alpha}^{\eps}$ the set of extremal points $\PBXgy$ that satisfy the inequality in  \eqref{eq:alpha cond} with equality.
    For $\alpha<\infty$, obtaining the extreme points in \eqref{eq:alpha cond} w.r.t $\PBXgy$ is not possible to the best of our knowledge. 
    Next, we use properties of $\alpha$-lift  to estimate some of these extreme points.

    According to Remark \ref{remark:non-decreasing}, for an extremal point $\mathbf{V} \in \mathcal{E}_{\alpha}^{\eps}$,  $\mathbf{W}=\PBSgX\mathbf{V}$  and any $\alpha'>\alpha$, we have  
    \begin{equation}\label{eq:nondec inequa}
        \ellalpha{\mathbf{W}}{\PBS}    \leq \ell_{\alpha'}\left(\mathbf{W}\|{\PBS}\right) \leq \ell_{\infty}\left(\mathbf{W}\|{\PBS}\right).
    \end{equation}  
    Using \eqref{eq:polytope}, for \(\alpha = \infty\) and any \(\eps'\), we obtain the extremal points \(\mathcal{E}^{\eps'}_{\infty}\) as the vertices of \(\Pi_{\eps'}\). The relations~\eqref{eq:nondec inequa} and~\eqref{eq:polytope} together prescribe a heuristic method for estimating some (but not all)
    of the extremal points in \(\mathcal{E}_{\alpha}^{\eps}\) for \(\alpha < \infty\) and \(\eps < \eps'\). Specifically, for \(\eps' > \eps\), we find vectors \(\mathbf{V} \in \mathcal{E}^{\eps'}_{\infty}\) that result in \(\ellalpha{\PBSgy}{\PBS}\) being very close to \(\e^{\eps}\). In the next subsection, we propose a heuristic algorithm to achieve this.

    \subsection{Heuristic algorithm for optimal utility estimation}
    
        \begin{algorithm}[t]
            \textbf{Input}: $\PBSX$, $\mathcal{A}=\{ \alpha_{1}, \ldots, \alpha_{|\mathcal{A}|} \}$, $\mathcal{B}=\{\epsilon_1, \ldots, \epsilon_{|\mathcal{B}|} \}$, $\mathcal{N}=\{ n_1, \ldots, n_{|\mathcal{B}|} \}$, $\mathcal{U}_{\alpha_{0}}^{\epsilon_j} = \emptyset$ for $1 \leq j \leq |\mathcal{B}|$, $\mathcal{U}_{\alpha_{i}}^{\epsilon_0} = \emptyset$ for $1 \leq i \leq |\mathcal{A}|$, $\delta$, $\epsilon_{|\mathcal{B}|+1}$.\\ 
          
            \textbf{Output}: $\Y_{i,j}$, $\PBXgY$, $\PBY$, and $I_{i,j}(X;Y)$ for $1 \leq i \leq |\mathcal{A}|$, $1 \leq j \leq |\mathcal{B}|$.\\
            
            \textbf{Initiate:}  Let $\mathcal{B}'_j=\{\eps_{j}+\frac{k(\eps_{j+1}-\eps_{j})}{n_{j}},0 \leq k \leq n_{j}-1\}$ for $1\leq j \leq |\mathcal{B}|$, $\mathcal{B}' = \cup_{j=1}^{|\mathcal{B}|} \mathcal{B}'_j$, and for $1\leq i\leq |\mathcal{A}|$, $1\leq j \leq |\mathcal{B}|$, let $\mathcal{F}^{\eps_{j}}_{\alpha_{i}}=\{\PBXgy \text{~in~}\eqref{eq:output Pxgy}\}$.\\
            Obtain $\mathcal{E}^{\eps'}_{\infty}$ for all $\eps' \in \mathcal{B}'$. \\
            \For{$i=1:|\mathcal{A}|$}{
            \For{$j=1:|\mathcal{B}|$}{
                $\mathcal{F}_{\alpha_{i}}^{\eps_{j}}\leftarrow 
                \mathcal{F}_{\alpha_{i}}^{\eps_{j}}
                \cup\mathcal{U}_{\alpha_{i-1}}^{\eps_{j}}
                \cup\mathcal{U}_{\alpha_{i}}^{\eps_{j-1}} $\\
                \For{$k=j:|\mathcal{B}|$}{
                    \begin{align*}\hspace{-10pt}
                       \nonumber \mathcal{F}_{\alpha_{i}}^{\eps_{j}}\leftarrow
                        \mathcal{F}_{\alpha_{i}}^{\eps_{j}}&
                        \cup \{\mathbf{V}\in\mathcal{E}^{\eps'}_{\infty}:\eps' \in \mathcal{B}'_k\\
                        &(1-\delta)\e^{\eps_{j}}\leq \ellalpha{\PBSgX\mathbf{V}}{\PBS} \leq\e^{\eps_{j}} \};
                    \end{align*}
                }
                Let  $\mathcal{V}_{\eps_{j}} = \mathcal{F}_{\alpha_{i}}^{\eps_{j}}$ and solve \eqref{eq:Py optimize}.
                Use \eqref{eq:output Y}-\eqref{eq:output Pxgy} to determine $\Y_{i}^{j}$, $\PBXgY$, $\PBY$, $I_{i}^{j}(X;Y)$. Let $\mathcal{U}_{\alpha_{i}}^{\eps_{j}}= \{ \PBXgy s\text{~obtained for~} (\alpha_{i},\eps_{j}) \}$;
            }
            } 
            \caption{$I(X;Y)$ estimation  given $\alpha$-lift privacy}\label{alg:Algorithm}
        \end{algorithm}

        \begin{figure*}[th]
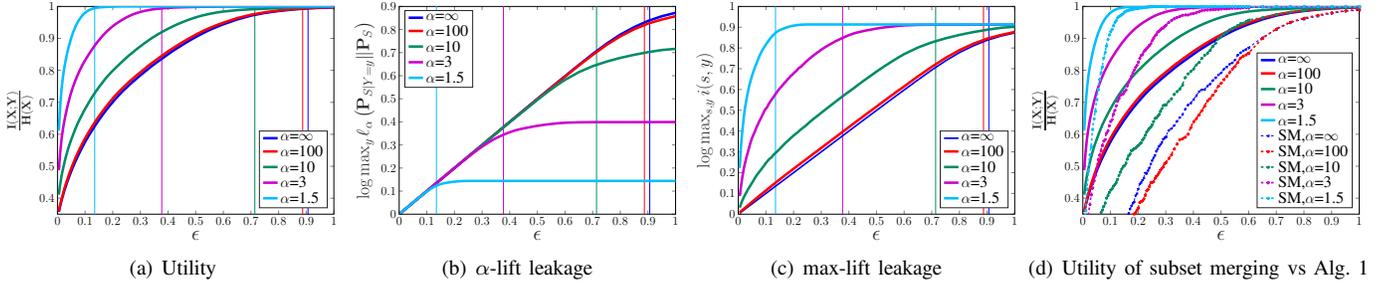

            \centering   
            \subfigure[\label{fig:utility} Utility]{\scalebox{0.32}{\input{figure1.tex}}}
            \subfigure[\label{fig:alpha leak} $\alpha$-lift leakage]{\scalebox{0.32}{\input{figure2.tex}}}
            \subfigure[\label{fig:maxlift} max-lift leakage]{\scalebox{0.32}{\input{figure3.tex}}}
            \subfigure[\label{fig:subset} Utility of subset merging vs Alg.~\ref{alg:Algorithm}]{\scalebox{0.32}{\input{figure4.tex}}}
            \caption{\label{fig:PUT} Privacy-utility tradeoff for $\alpha$-lift as privacy measure and normalized mutual information as utility measure. } 
        \end{figure*}

        We propose Algorithm~\ref{alg:Algorithm}, which is a heuristic approach to find solutions for the problem in~\eqref{eq:alpha optim}-\eqref{eq:alpha cond}. 
        Algorithm~\ref{alg:Algorithm} takes \(\PBSX\) and two ordered sets \(\mathcal{A}=\{\alpha_{1}, \ldots, \alpha_{|\mathcal{A}|}\}\), and \(\mathcal{B}=\{\eps_1, \ldots, \eps_{|\mathcal{B}|}\}\) as its main input, where \(\alpha_1 > \alpha_2>\cdots>\alpha_{|\mathcal{A}|}\) and \(\eps_1<\eps_2<\cdots<\eps_{|\mathcal{B}|}\). 
        Other auxiliary parameters will be explained shortly.  The output is a feasible solution for \(\PBXgY, \PBY\) and the corresponding output alphabet $\Y_{i,j}$ and utility \(I_{i,j}(X;Y)\) for each pair of \((\alpha_{i}, \eps_{j})\).
        
        The utility value is generally expected to increase when $\eps$ increases and when $\alpha$ decreases. However, due to the heuristic nature of Algorithm \ref{alg:Algorithm}, one may obtain a slightly lower utility for $(\alpha_{i}, \eps_{j})$ compared to the utility of $(\alpha_{i'}, \eps_{j'})$ where  \(\alpha_i < \alpha_{i'}\) or $\eps_{j}>\eps_{j'}$. To address this, we propose considering multiple \(\eps\) and \(\alpha\) values as follows.
        
        1) The algorithm initializes $\mathcal{F}_{\alpha_{i}}^{\eps_{j}}$ by $\PXgy$s in \eqref{eq:output Pxgy},  for $\alpha=\infty$ and each $\eps_{j} \in \mathcal{B}$ (line 3).  These vectors are feasible but not extremal for $(\alpha_{i},\eps_{j})$. Nevertheless, they guarantee that if a solution is optimally achievable for $\alpha=\infty$ and $\eps_{j}$ it is also achievable for any $(\alpha_{i}<\infty,\eps_{j})$.
        
        2) Recalling Remark \ref{remark:non-decreasing}, we note that the solution $\PBXgY$ for $(\alpha_{i},\eps_{j})$ is also feasible for $\alpha_{i'}<\alpha_{i}$ or $\eps_{j'} > \eps_{j}$. Therefore in line 7, for each pair of $(\alpha_{i},\eps_{j})$, we make a union of the feasible set $\mathcal{F}_{\alpha_{i}}^{\eps_{j}}$ and the previous solutions for $(\alpha_{i-1},\eps_{j})$ and $(\alpha_{i},\eps_{j-1})$ saved in $\mathcal{U}_{\alpha_{i-1}}^{\eps_{j}}$ and $\mathcal{U}_{\alpha_{i}}^{\eps_{j-1}}$, respectively. This ensures that if a solution is achievable for $\alpha_{i-1}>\alpha_{i}$ or $\eps_{j-1}<\eps_{j}$, then it is also achievable for $(\alpha_{i},\eps_{j})$. Note the input $\mathcal{U}_{\alpha_{0}}^{\eps_{j}}=\mathcal{U}_{\alpha_{i}}^{\eps_{0}}=\emptyset$ for all $i,j$ makes line 7 valid even for $i, j = 1$.
        
        Once $\mathcal{F}_{\alpha_{i}}^{\eps_{j}}$ is set as described above, in lines 8-11, we add extra points to it as follows. The values in set $\mathcal{B'}_j$ are set to $\eps' = \eps_{j}+\frac{k(\eps_{j+1}-\eps_{j})}{n_{j}}, 0 \leq k \leq n_{j}-1$, during initialization. For each $\eps' \in \mathcal{B'}_j$, we compute  $\mathcal{E}^{\eps'}_{\infty}$, the vertices of $\Pi_{\eps'}$ according to \eqref{eq:polytope} and add those elements $\mathbf{V} \in \mathcal{E}^{\eps'}_{\infty}$ such that their corresponding $\alpha$-lift is smaller than or equal to $e^{\eps_{j}}$, but bigger than or equal to $(1-\delta) e^{\eps_{j}}$ for some small $\delta$. 
        For the largest value $\eps_{|\mathcal{B}|}$, we use the auxiliary input $\eps_{|\mathcal{B}|+1}>\eps_{|\mathcal{B}|}$ and interpolate $n_{|\mathcal{B}|}$ points between them to estimate $\mathcal{E}_{\alpha_i}^{\eps_{|\mathcal{B}|}}$.

        Finally, with $\mathcal{F}_{\alpha_{i}}^{\eps_{j}}$ in hand, in line 12 we solve \eqref{eq:Py optimize} for $(\alpha_{i},\eps_{j})$ and calculate the utility by the derived $\PBYgX$ and $\PBY$ from \eqref{eq:output Y}-\eqref{eq:output Pxgy}. The answer is saved as $\mathcal{U}_{\alpha_{i}}^{\eps_{j}}$ for the following steps in the loops. In the next section, we examine the efficacy of this method via numerical evaluation.

     
\section{Numerical Results}
    
    We evaluate Algorithm~\ref{alg:Algorithm} via numerical simulations. We have generated $100$ distributions $\PBSX$ with $|\Sen|=6$ and $|\X|=10$, and the presented results are averaged over these distributions where 
    $\mathcal{A}=\{\infty, 100, 10, 3,  1.5\}$ and $\mathcal{B}=\{0.005, 0.01, 0.015, \cdots, 0.95\}$.
    We set $\delta=10^{-2}$, leading to a $1\%$ estimation error for extreme vectors.
    We also set $\eps_{|\mathcal{B}|+1}=1$ and $n_j = 3$ for all $j$. The averaged value of $\eps_{\alpha_{i}}^{\text{max}}$ is shown in Figure \ref{fig:PUT} by vertical lines with the corresponding color of $\alpha_{i}$.
    
    Figure \ref{fig:PUT} shows the results versus the $\alpha$-lift privacy budget $\eps$, where the normalized utility, $\frac{I(X:Y)}{H(X)}$, is shown in Fig.  \ref{fig:utility}, $\alpha$-lift leakage, $\max_{y}\log\ellalpha{\PBSgy}{\PBS}$, in Fig. \ref{fig:alpha leak}, and max-lift leakage, $\max_{y,s}i(s,y)$, in Fig. \ref{fig:maxlift}, respectively.
    These numerical results show a general trend that can be used to interpret the operational significance of choosing suitable values of $\alpha$ for PUT optimization. 
    Fig. \ref{fig:utility} shows that the utility is decreasing in  $\alpha$, where for a fixed $\eps$, a smaller value of $\alpha$  results in higher utility.
    One can observe that the utility and privacy leakage values are almost the same for $\alpha=\infty$ and $\alpha=100$.
    This can be explained by referring to the properties of $\alpha$-lift: for large $\alpha$, larger lift values become dominant and, after some point (e.g., in our simulations $\alpha=100$), only the max-lift is effective, which is equal to $\alpha$-lift for $\alpha=\infty$. 
    
    In Fig. \ref{fig:alpha leak}, we observe that the increment rate of $\alpha$-leakage decreases after a value of $\eps$ depending on the value of $\alpha$.
    For instance, consider $\alpha=3$ where $\alpha$-leakage increment rate decreases after $\eps\approx 0.37$ and remains almost constant for $\eps>0.55$. 
    The reason is that the values of $0.37$ is the mean value of $\eps^{max}_{3}$, after which there is no extremal vector to satisfy $\ellalpha{\PBSgX\PBXgy}{\PBS}=\e^{\eps}$. 
    If we compare Fig.~\ref{fig:utility} with Fig.~\ref{fig:alpha leak} we see that the utility values are very close to $1$ after the corresponding $\eps^{max}_{\alpha_{i}}$, which means $\PBXgY$ is the unity matrix and there is no privacy.
    Figure. \ref{fig:maxlift} depicts max-lift leakage for each $\alpha$ and confirms  that while $\alpha$-lift is restricted by $\e^{\eps}$, max-lift leakage could be much larger than this value. This is a natural property of $\alpha$-lift which is a more relaxed measure compared to the max-lift.
    In Fig. \ref{fig:subset}, we have compared the utility values in Fig. \ref{fig:utility} with the corresponding values given by the subset merging (SM) method in \cite{2023Onthelift}. Algorithm \ref{alg:Algorithm} enhances utility significantly, confirming its near optimality. 
    Moreover, subset merging does not maintain the properties of $\alpha$-lift privacy for $\alpha=100$ as its utility value is less than that of $\alpha=\infty.$ 
    Note that utility values for subset merging are very small for $\eps<0.1$ and large $\alpha$. These are not shown in order not to clutter the figure.
    
\newpage
\bibliographystyle{IEEEtran}
\bibliography{IEEEabrv.bib, BIB}
\end{document}